# Gouverner la standardisation par les changements d'arène.

## LE CAS DU XML

**Pablo Andres Diaz**
Etudiant-chercheur à l'Université de Lausanne (IEPI)

**François-Xavier Dudouet**
CR 1 CNRS en sociologie, Université Paris Dauphine (IRISES)

**Jean-Christophe Graz**
Professeur de science politique, Université de Lausanne (IEPI-CRII)

**Benjamin Nguyen**
Maître de conférences en informatique, Université de Versailles (PRISM)

**Antoine Vion**
Maître de conférences en science politique à l'Université de la Méditerranée (LEST)

# 1. Introduction

Malgré l'important développement des travaux sur la standardisation au cours des trente dernières années, tant en économie industrielle qu'en sociologie, peu d'études ont essayé de prendre en compte la question des structures de gouvernance dans le travail de standardisation.

Comme Mattli l'a relevé, "the literature on standards setting generally lacks a sustained theoretical argument to explain or assess *institutional* standards arrangements past or present" (Mattli, 2001, p. 331). La majeure partie des travaux sur ces questions est fortement imprégnée par des modèles de microéconomie industrielle ou de théorie des jeux. Dans ce papier, nous discuterons d'abord les limites de ces approches, pour mettre en avant la nécessité de perspectives d'économie politique internationale. Ces perspectives cherchent en effet à caractériser les changements récents dans la distribution ou la dévolution du pouvoir de standardisation. Un phénomène qui retient l'intérêt est évidemment dans ce domaine la multiplication des normes de consortium qui fait évoluer la coordination du travail vers des forums hybrides (Graz, 2006 ; De Nardis, 2008) ou vers des formes de gouvernance transnationale privée (Graz, Nölke, 2007 ; Debardeau, 2008 ; Dudouet, Nguyen, Vion, 2008). Ce constat d'un mouvement de privatisation, plus ou moins accusé selon les cas, fait converger des approches néo-schumpeteriennes centrées sur les stratégies de firmes (Dudouet, Mercier, Vion, 2006 ; Dudouet, Vion, 2009), et des approches néo-gramsciennes qui insistent davantage sur le sens des nouveaux rapports de force entre organisations publiques et privées (Debardeau, 2008), et les changements structurels dans la dévolution des pouvoirs, marqués par la privatisation et la transnationalisation, avec toutes les limites que cela comporte du point de vue du contrôle démocratique (Graz, Nölke, 2007).

Il reste néanmoins un enjeu méthodologique commun à toutes ces études, qui consiste à dynamiser l'analyse, en développant des cas empiriques qui permettent de prendre toute la mesure de ces changements et du rôle qu'y jouent les acteurs tout au long de la séquence de standardisation (Tamm-Hallströmm, 2001, 2004, 2008). Le domaine des TIC est ici particulièrement intéressant, puisque le travail de standardisation s'y déploie sur une à deux décennies pour une technologie donnée, comme l'illustrent les cas des standards de téléphonie mobile (Bekkers et al. ; Dudouet, Grémont, Vion, ; Funk, ) ou des standards de l'internet et du du web (De Nardis, 2008 ; Dudouet, Nguyen, Vion, 2008 ; Jacobs, 2009).

Dans ce papier, nous discutons les approches disponibles des structures de gouvernance de la standardisation (2) pour proposer de nouvelles hypothèses sur celle des langages informatiques (3). Nous développons ensuite le cas du langage XML et de ses applications pour proposer une analyse dynamique de cette gouvernance, en insistant sur la coordination qu'y assurent les firmes, et l'usage stratégique qu'elles font des arènes dans la poursuite de leurs objectifs (4). Nous plaidons ensuite pour le développement de telles analyses empiriques dynamiques afin d'assurer le rayonnement des perspectives d'économie politique internationale dans ce domaine (5).

## 2. Etudier la gouvernance de la standardisation : enjeux analytiques et modèles théoriques.

"Governance in a world where boundaries are largely in flux is being shaped and pursued in constellations of public and private actors that include states, international organizations, professional associations, expert groups, civil society groups and business corporations." (Djelic, Sahlin-Andersson, 2006). Dans le domaine de la standardisation, la question relative aux modalités de coordination de ces différents acteurs se pose de manière récurrente. Les études économiques apportent peu de réponses à ces questions, alors que la qualification des structures de gouvernance de la standardisation est un enjeu important pour l'analyse des changements en cours (Graz, 2006 ; Dudouet, Vion, 2009).

### 2.1. Limites de la modélisation économique des structures de gouvernance de la standardisation

Chiao, Lerner, et Tirole (2007) ont développé l'une des modélisations les plus abouties concernant la relation entre les politiques suivies par les organisations de standardisation et les stratégies des firmes. Très justement, les auteurs constatent que la littérature économique sur le sujet est relativement limitée, à la fois en termes de nombre d'articles et de portée analytique. Dans le modèle proposé par Farrell et Saloner (1988), deux firmes peuvent choisir entre deux technologies incompatibles, et peuvent le faire en échangeant de façon régulière au sein d'une organisation de standardisation, ou à travers une compétition sur le marché, ou selon une stratégie hybridant les deux types de trajets. Comme le relèvent Chiao, Lerner et Tirole, l'organisation de standardisation dans leur modèle est très pauvre, puisqu'elle est seulement un lieu de négociation, mais n'est pas considérée comme une institution apte à imposer des règles de décision ou des *prerequisits,* comme la déclaration préalable des brevets (comme souvent dans les organisations définissant des *sponsored standards*) ou le renoncement à ces droits (comme souvent pour les *open standards*). Au-delà de ce problème théorique essentiel se pose un problème méthodologique : celui du choix de n'analyser qu'une seule structure. Dans un contexte où la plupart du temps les firmes changent de structure, cela revient à pronostiquer l'issue d'une partie d'échecs en analysant les mouvements de pion sur une aile en début de partie, et sans se préoccuper du reste. Cela peut avoir une certaine portée esthétique, mais nous en dit peu sur la dynamique effective de la partie qui est jouée.

Le mérite du papier publié dans le *Rand Journal of Economics* (Chiao, Lerner, Tirole, 2007) est de prendre au sérieux la diversité des contraintes posées par les institutions et des alternatives qui s'offrent aux firmes, pour élaborer un modèle prédictif susceptible d'interroger l'attractivité des politiques des organisations pour les promoteurs d'une solution technologique. Le modèle prend ainsi en compte les degrés de contrainte posés par les concessions préalables demandées par les organisations, les déterminants de l'ouverture des informations réciproques, et les positionnements dans une compétition entre organisations.

Malgré la très grande pertinence du cadre analytique et la richesse des données, deux critiques principales peuvent être adressées à cette étude. La première est liée aux limites des méthodes de bouclage des trajets décisionnels de la théorie des jeux relevées par Pierre Livet: « Pour assurer ce bouclage, en théorie des jeux, on ne retiendra que les chemins qui mènent de manière causale continue d'une étape à une autre. Nous prenons par exemple le point de vue du joueur de l'étape e dans le noeud n, et nous éliminons ceux des chemins qui passant par e mènent à des résultats pour le joueur de l'étape e-1 qui sont inférieurs à ceux donnés pour ce joueur précédent par d'autres chemins. En effet si le joueur précédent a anticipé que le joueur suivant orienterait le jeu vers un chemin qui est moins bon pour lui que le moins bon des résultats des autres branches, il aura bloqué le chemin causal vers le noeud où le joueur suivant est tenté de faire ce choix, et le joueur suivant faisant ce raisonnement contrefactuel aura éliminé ce chemin. On élimine ainsi peu à peu les chemins dans l'arbre du jeu pour n'en conserver qu'un seul, celui qui n'est pas ainsi exclu » (Livert, 2007, p. 2). Livet montre bien que tout jeu social se caractérise au contraire par ce qu'il nomme la transmutation des évaluations, qui met en jeu de multiples états émotionnels dans la comparaison des choix, et ouvre des possibles très complexes de rétro-évaluations ou de réévaluation de certaines solutions, qui conduisent à des trajets beaucoup plus obliques que les trajets linéaires proposés dans ce type de modèle pour caractériser les choix des firmes. La deuxième critique, plus empirique, tient au caractère trop statique de l'analyse du processus global, comme le relèvent finalement les auteurs dans leur conclusion : « One of the most intriguing of these has to do with the *dynamics* of certification. Lerner and Tirole (2008) and the extensions discussed here present a static model in which a one-time decision is made. In the real world, SSOs and sponsors may employ more complex strategies: for instance, a sponsor may reapply to an SSO after its initial application is rejected. (Farhi, Lerner, and Tirole, 2006 present a theoretical look at these issues.)

Understanding the dynamics of the certification process represents an important empirical challenge. » (Chiao, Lerner, Tirole, 2007, p. 927).

C'est pourquoi nous privilégions au contraire cette entrée dynamique, en essayant de caractériser le processus à partir de tableaux de pensée de moyenne portée construits de façon idéaltypique, selon une méthode weberienne. Pour parler comme Gérard-Varet et Passeron (1997), nous préférerons l'enquête au modèle.

## 2.2. Gouvernance en réseau versus gouvernance transnationale privée

Dans le domaine de l'analyse des nouvelles technologies, la notion de gouvernance en réseau (*network governance*, puis *networked governance*) a été développée ces dernières années (Borgatti & Foster, 2003 ; Olsen, 2005 ; Lazer, 2007 ; Levinson, 2008). Cette notion, développée d'abord par Jones, Hesterly et Borgatti (Jones et al., 1997), s'attache à mettre en avant un certain nombre de conditions d'échange par lesquelles émerge une structure de gouvernance particulière, caractérisée par un accès restreint à l'échange, l'existence d'une culture commune, la prédominance des logiques réputationnelles et des mécanismes de sanction des comportements non conformes à la culture commune. Ces conditions d'échange sont des conditions relativement bien connues des économistes et sociologues de l'innovation : l'incertitude sur la demande, la spécificité des actifs humains (notamment en termes de management de la connaissance), la complexité des tâches initiées, et la fréquence des interactions qu'elles supposent dans la conduite des projets.

Sans s'attacher à définir l'ensemble de la gouvernance de la standardisation, les travaux relatifs à l'émergence de l'*open source* font largement écho à cette conception de la gouvernance en réseau. Contre la logique des logiciels propriétaires protégés par le secret des codes sources, des réseaux de chercheurs plus ou moins ouverts ont développé une culture commune de l'innovation partagée (Von Hippel, 2001, 2005 ; Weber, 2004), qui brouille les frontières classiques entre producteurs et consommateurs (De Landa, 2001, O'Mahoney, 2002), et fait émerger de nouvelles logiques réputationnelles – la communauté Linux étant certainement le cas le plus célèbre en la matière. S'il ne faut pas surestimer la capacité des particuliers motivés (*geeks*) à mener des projets durables en *open source* (David, Rullani, 2008), le succès de ce type de gouvernance a sans doute tenu au départ au fait qu'il développait une logique de *happy few* issus de mondes sociaux très divers (centres de recherche publics, ONG, ingénieurs du privé, apprentis informaticiens, autodidactes, etc.), capables de répondre par eux-mêmes à de nouveaux besoins de programmation (Von Hippel, Von Krogh, 2003 ; Gallaway, Kinnear, 2004) et d'apprendre ensemble (Lakhani, Von Hippel, 2005 ; Jullien, Zimmermann, 2006). Si peu de spécialistes parlent de gouvernance en réseau à ce sujet, on peut développer l'idée que les processus de la standardisation des innovations *open source* correspondent dans un certain nombre de cas à ce type de gouvernance.

D'un autre côté, d'autres travaux relatifs à la standardisation d'Internet ont plus nettement inscrit celle-ci dans le modèle de la gouvernance transnationale privée définie par Graz et Nölke (2007) comme "the ability of non-state actors to cooperate across borders in order to establish rules and standards of behaviour accepted as legitimate by agents not involved in their definition. Non-state actors not only formulate norms, but often also have a key role in their enforcement. Accordingly, the current privatization of rule-making and enforcement goes much further than traditional lobbying in allowing private actors an active role in regulation itself" (Graz, Nolke, 2007). A partir des développements de leur ouvrage, nous pouvons comparer ce modèle avec celui de la gouvernance en réseau en utilisant les mêmes catégories analytiques que Jones et al. (1997). Dans le modèle de la gouvernance transnationale privée, les conditions d'échange sont beaucoup plus asymétriques, marquées par la transformation rapide des marchés, la domination oligopolistique, la dépendance des acteurs économiques vis-à-vis de chaînes de valeur globales, et la lourdeur des mécanismes de coordination. Les formes d'encastrement de la gouvernance sont donc beaucoup plus liées aux ressources des grandes firmes multinationales, qu'il s'agisse de ressources financières (*fund rising*), procédurales (multiplicité des procédures formelles ou informelles), relationnelles (loyauté ou rapport de force), ou symboliques (image de marque, crédibilité, etc.). C'est principalement sur la circulation de ces ressources que s'appuie la structuration de la gouvernance transnationale privée, qui ne suppose pas nécessairement un accès restreint à l'échange. Dans le domaine de la standardisation, il est fréquent que s'imposent des règles de 'pay to play' qui laissent le jeu ouvert, mais qui, comme dans les grands tournois de poker, favorisent ceux qui ont au départ la plus grande profondeur de tapis.

Nous arrivons ainsi au tableau suivant :

|  | **Gouvernance en réseau** (d'après Borgatti et al., 1997) | **Gouvernance transnationale privée** (d'après Graz et Nölke, 2007) |
|---|---|---|
| **Conditions d'échange et de coordination** | • Incertitude sur la demande<br><br>• Complexité des tâches<br><br>• Spécificité des actifs humains<br><br>• Fréquence des interactions | • Transformation de l'organisation des marchés<br><br>• Lourdeur des mécanismes de coordination<br><br>• Compétition limitée (oligopolistique)<br><br>• Dépendance vis-à-vis de chaînes de valeur globales |
| **Formes d'encastrement de la gouvernance** | • Accès restreint à l'échange<br><br>• Réputation et légitimité professionnelle<br><br>• Sanction collective des comportements non conformes<br><br>• Communauté épistémique ou culture technique partagée | • Procédures formelles et informelles<br><br>• Crédibilité de la firme<br><br>• Comportements loyaux ou rapports de force<br><br>• Capacités financières et pouvoir de lever des fonds |

**Tableau 1. Comparaison des modèles analytiques de la gouvernance en réseau et de la gouvernance transnationale privée, d'après Borgatti et al., 1997, et Graz et Nölke, 2007.**

## 3. Quelle gouvernance dans le domaine de la standardisation informatique ?

Une fois définis ces modèles de gouvernance, nous pouvons nous interroger sur les formes institutionnelles et les modes de coordination qui prévalent dans le domaine de la standardisation informatique. Dans ce domaine, de nombreuses études ont insisté sur la standardisation *de facto* exercée par certaines grandes firmes comme IBM ou Microsoft. Mais la multiplication des législations anti-trust et le renforcement des politiques de concurrence rendent cette dynamique de standardisation de plus en plus coûteuse pour les firmes. Il nous semble donc important de reconsidérer le paysage organisationnel de la standardisation et de mettre en avant la diversité des règles en vigueur, avant d'étudier l'interdépendance entre les organisations et de définir de nouvelles hypothèses sur les modalités empiriques de la coordination du travail de standardisation.

### 3.1. Typologie des arènes de normalisation dans la standardisation informatique

Si nous considérons les arènes de standardisation informatique, nous identifions trois types de structures : les organisations internationales à vocation générale, les organisations transnationales spécialisées, et les consortiums industriels régionaux ou globaux.

Nous nommons **organisations internationales** toutes les organisations structurées sur la base de représentations nationales, qu'elles soient intergouvernementales, privées ou hybrides.

La plus ancienne de ces organisations est l'Union Internationale des Télécommunications, créée à la fin du XIXe siècle pour standardiser le télégraphe, et qui intervient aujourd'hui dans la standardisation informatique pour tout ce qui concerne les protocoles de l'Internet ou des dernières générations de téléphonie mobile. L'UIT intervient le plus souvent ici en dernier ressort, pour asseoir la légitimité et l'universalité de standards déjà consacrés par d'autres organisations. Si l'UIT fonctionne à la base de manière intergouvernementale, il faut noter l'ouverture croissante à la représentation des entreprises dans les processus de décision de l'organisation.

Une organisation comme ISO peut aussi être identifiée comme organisation internationale, dans la mesure où, même si elle mixe des représentations par des agences publiques et privées, toutes ces représentations sont uniques et définies sur une base nationale, de sorte qu'il y a toujours un encastrement national des travaux et des options de vote, avec tous les problèmes classiques d'asymétrie que cela pose, comme nous le verrons. Le règlement de l'ISO stipule que chaque comité national doit spécifier s'il prétend :

> - participer activement au travaux d'un comité technique, avec l'obligation de voter sur toutes les questions formellement soumises au vote ainsi que de participer activement aux réunions. (**P-member**) ou
>
> - suivre les travaux d'un comité technique en tant qu'observateur, recevoir les documents tout en ayant le droit de faire des commentaires ainsi que d'assister aux réunions (**O-members**)

Un comité national peut également choisir de n'être ni P-member ni O-member d'un comité technique. Dans quel cas, il n'a ni les obligations ni les droits précédemment évoqués. Cela dit, tous les membres nationaux peuvent voter, indépendamment de leur statut, en ce qui concerne les *enquiry drafts* et les *final draft International Standards*. En ce qui concerne le vote à proprement parler, un projet devient une norme lorsqu'il est approuvé par au moins 66% des P-members du comité technique aillant travaillé dessus et n'est pas rejeté par plus de 25 % de tous les votants.

On peut aussi associer à ce premier type des organisations comme l'IEC.

Le deuxième type est celui des **organisations transnationales spécialisées**, qui sont généralement issues d'initiatives pionnières dans le développement d'une nouvelle technologie. Dans le domaine informatique, on peut ainsi mentionner le consortium Unicode, qui a joué un rôle moteur dans le développement des normes typographiques et des formats de représentation, l'IETF (Internet Engineering Task Force), qui a initié la standardisation des protocoles IP et des noms de domaine, ou encore le W3C, qui a initié la standardisation des langages du web, de l'HTML au SQL, jusqu'au XML que nous étudierons ici. Notons qu'interviennent aussi dans la standardisation informatique, mais dans une moindre mesure, des organisations du même type structurées sur d'autres enjeux comme le développement de la téléphonie mobile de troisième et quatrième génération (ETSI, 3GPP, notamment).

Enfin, il nous faut mentionner la très grande diversité des **consortiums industriels régionaux ou globaux**, qui définissent des standards dans le cadre de groupements d'entreprises dotés de règles de vote tout aussi contraignantes que les organisations précédentes, mais qui ont des politiques variables en matière de prérequis. Dans cette étude, nous mentionnerons notamment les consortiums OASIS et ECMA, qui sont les deux plus imporatnts en informatique, mais il faut garder à l'esprit que de nombreux autres consortiums de ce type interviennent sur les débats informatiques dans le cadre de la standardisation de l'Internet ou du web : consortium GSM, Wimax Forum, open Mobile Alliance, etc.

Toutes ces organisations sont en interdépendance. Ces interdépendances échappent largement au contrôle des Etats, ou à une logique de régulation globale, et on peut donc parler ici de gouvernance. La question qui se pose est donc de savoir comment s'effectue la coordination entre les arènes, comme se structure cette gouvernance de la standardisation informatique. Nous proposons ici de nouvelles hypothèses.

### 3.2 Nouvelles hypothèses sur la capacité de coordination des multinationales du secteur

Dans ce papier, nous sommes partis de l'hypothèse selon laquelle les plus grandes firmes multinationales assurent la majeure partie de la coordination du travail de standardisation informatique, malgré les logiques émergentes de gouvernance en réseau de *l'open source*. Nous constatons en effet que les firmes contrôlent le travail de standardisation à l'intérieur des organisations, qu'il s'agisse du W3C (Dudouet, Nguyen, Vion, 2008) ou de l'IETF (Jacobs, 2009). Même si le contrôle des votes à ISO est plus complexe, il faut noter d'emblée que les règles d'ISO offrent aux consortiums industriels des entrées privilégiées, comme la procédure de Public Available Specification pour OASIS, qui lui permet de faire voter rapidement toute nouvelle application ouverte, ou la procédure de Fast Track pour ECMA, qui accélère les débats lorsqu'un besoin urgent se fait jour.

Il résulte de cette configuration une forte asymétrie, puisque les agences qui proposent et votent à ISO ne sont pas présentes dans les autres organisations, ce qui les amène à une moindre spécialisation, alors que les firmes sont présentes partout à partir du moment où elles 'pay to play'. Dans ce papier, nous allons monter que les firmes jouent de cette asymétrie pour prendre le leadership dans le travail de standardisation, en faisant un usage stratégique des arènes. Le cas du XML est exemplaire de ce type de stratégies.

## 4. L'usage stratégique des changements d'arène : le cas du XML

L'univers de la standardisation internationale est aujourd'hui parsemé d'agences et de forums qui tous prétendent intervenir à un titre ou un autre dans la production de nouveaux standards. Il existe de fait une concurrence entre les différentes arènes que les entreprises entretiennent d'autant plus volontiers qu'elle peut servir leurs intérêts. En effet, le positionnement des arènes, leur histoire, l'autorité qu'il leur est reconnu, sont autant de ressources que les entreprises cherchent à capter dans un sens conforme à leurs intérêts. De là, le développement d'usages stratégiques des arènes de standardisation consistant à investir ou désinvestir telle ou telle d'entre elles en fonction des coups à jouer. Braithwaite et Drahos (2000) ont nommé stratégies de *forum shifting* les stratégies menées par les Etats les plus puissants consistant à passer d'une organisation internationale à une autre pour imposer leur agenda et leur trend de réformes. Nous observons le même type de stratégies du côté des entreprises multinationales à l'égard des organisations de standardisation. Par exemple le W3C et l'ISO offrent des ressources très différentes mais néanmoins complémentaires pour les firmes qui savent opportunément s'en saisir. Le W3C ne produit pas vraiment des standards mais des recommandations qui n'ont aucune valeur contraignante. En revanche son autorité en matière de technologie du Web est telle que ses recommandations sont des quasi standards. L'ISO de son coté n'a pas de compétence reconnue en matière de technologies du Web, mais est en revanche la seule agence à visée universelle en mesure de produire des standards relativement contraignants. Composé des agences nationales de normalisation, toute norme adoptée par l'ISO devient *de facto* une norme nationale pour l'ensemble du secteur d'activité considéré.

Le maquis de la standardisation internationale peut, à qui sait en saisir les opportunités, devenir une source de profits considérables. Mais on aurait tort d'imaginer que les firmes sont en mesure de mettre en œuvre des stratégies claires et précises. Par ses rapports étroits avec l'innovation, la multiplicité croissantes des intervenants, les retournements d'alliance et l'évolution des configurations d'expertises, la standardisation demeure fondamentalement soumise à l'incertitude. La possibilité de jouer des coups gagnants dépend finalement des opportunités qui éclaircissent momentanément le jeu et permettent d'avoir une vision plus précise des rapports de force. Le cas du XML en est une bonne illustration.

## 4.1. Les changements d'arène comme ressource dans la compétition

De part ses origines, l'univers du Web est traditionnellement réfractaire à toute idée de brevet. Les efforts déployés pour promouvoir des technologies libres de droit est la marque distinctive et continue de ce milieu. Le W3C, sans doute l'agence la plus influente dans la normalisation du Web, n'échappe pas à cette tendance au point d'avoir érigé règle de fonctionnement le principe de l'open source. Les firmes qui participent aux travaux du W3C sont contraintes de renoncer à tout droit de propriété sur les normes élaborées en son sein. Dans ce cas de figure peu de chance de rencontrer, comme dans la téléphonie, un processus de normalisation entièrement contrôlé par les firmes où tout du moins recouvrant des enjeux économiques cruciaux à même de décider de la survie ou de la mort de grands groupes industriels. Pourtant la simple lecture de la liste des membres du W3C fait apparaître un très grand nombre d'entreprises informatiques parmi les plus importantes au monde dont celles qui comme Microsoft étaient au début des années 2000 les moins susceptibles de défendre l'open source. On pourrait penser en première approche que leur présence n'est pas forcément active et qu'elles n'investissent le W3C que pour mieux surveiller l'évolution des recommandations en discussion. Cet investissement est d'autant moins coûteux que l'essentiel du travail s'effectue dans ces nouvelles arènes virtuelles de la normalisation que sont les listes de discussion. En effet, l'activité des experts au sein du W3C consiste principalement à argumenter et négocier sur des listes de diffusion au sein desquelles les recommandations techniques sont débattues.

### *Le contrôle du travail du W3C par les corporate rulers*

Les analyses menées sur huit listes de discussion du W3C participant à l'élaboration du XML[1] montrent au contraire une forte activité des experts issus des firmes d'informatique. Le Graphe 2 représente l'activisme des experts (nombre de message envoyé[2]) sur les différentes listes entre 2000 et 2006. Les experts eux-mêmes ont été substitués par leurs institutions de rattachement afin de mieux rendre compte de l'investissement concret des différents types d'institution. La configuration d'ensemble est formée par quatre type d'acteur : les entreprises d'informatique, notamment les leaders mondiaux, les centres de recherche publics dominés par les grandes universités anglo-saxonnes, des organisations non gouvernementales et enfin quelques acteurs indépendants, ne représentant qu'eux mêmes. On peut observer un très fort activisme des grands noms de l'informatique comme Microsoft, IBM, Oracle ou Sun Microsystem. Les centres de recherche publics ou les organisations non gouvernementales ne sont bien sûr pas inactives, mais elles sont loin de dominer les débats.

---

[1] Ces analyses ont été menées dans le cadre du projet Webstand (ANR STIC 2006-2009), auquel ont participé Benjamin Nguyen, François-Xavier Dudouet, Antoine Vion, Ioana Manolescu, Dario Colazzo et Pierre Senellart.
[2] N'ont été pris en compte pour ce calcul que les individus ayant posté au moins 20 messages.

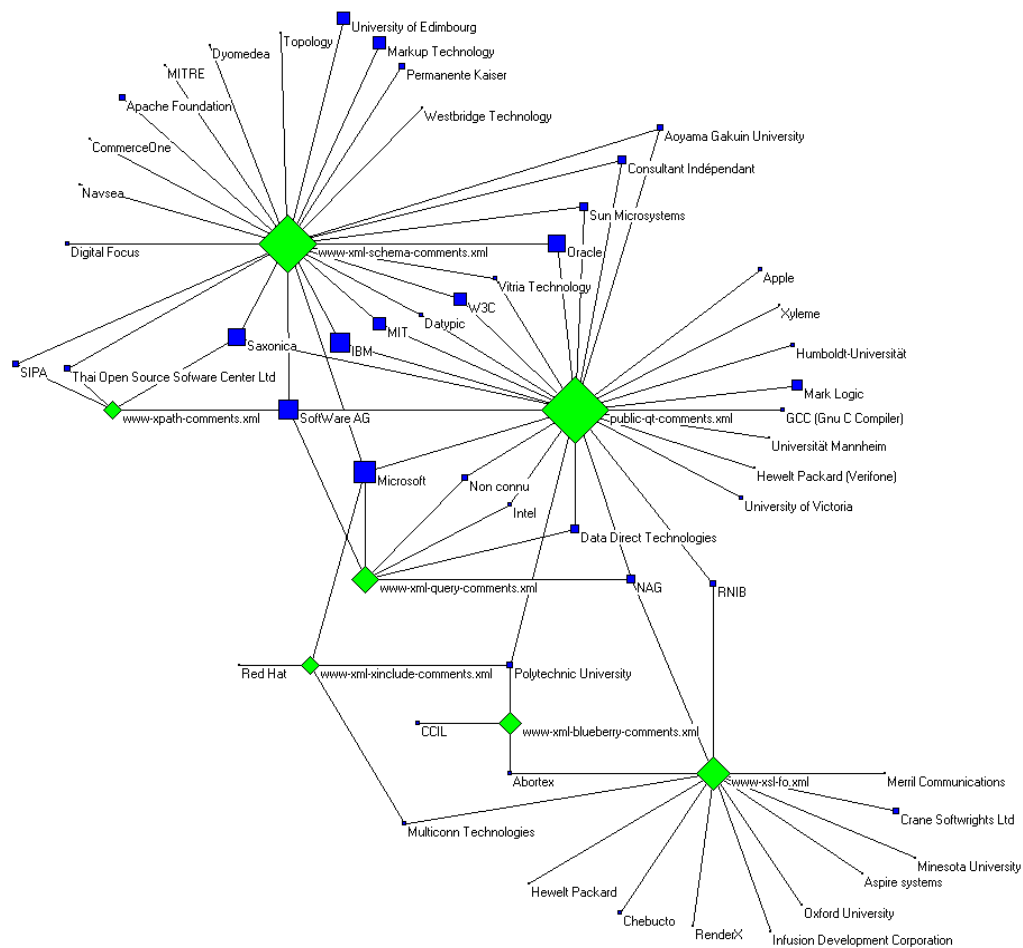

**Graphe 1. Activisme des institutions au sein des listes diffusion publiques concernant les standards XML (plus particulièrement X Query)**

**Losange = liste de discussion / Carré = Institution**

**La taille des nœuds représente le nombre de message envoyé (institutions) ou reçus (listes de discussion)**

Un examen des auteurs des préconisations issues des travaux de ces huit listes de discussion montre que celles-ci ont essentiellement été produites par les experts représentants les firmes. Le graphe 2 représente le réseau des co-auteurs des textes sur le XML liés à ces huit listes. On remarque une configuration nettement dominée par les experts des firmes notamment d'At&T, IBM, Oracle et dans une moindre mesure Microsoft. Les centre de recherche public et les ONG sont clairement repoussés à la périphérie du réseau. Cela signifie deux choses : d'une part, ils écrivent moins de textes que leurs homologues des entreprises et d'autre part, ils sont moins souvent engagés dans des processus de co-écriture.

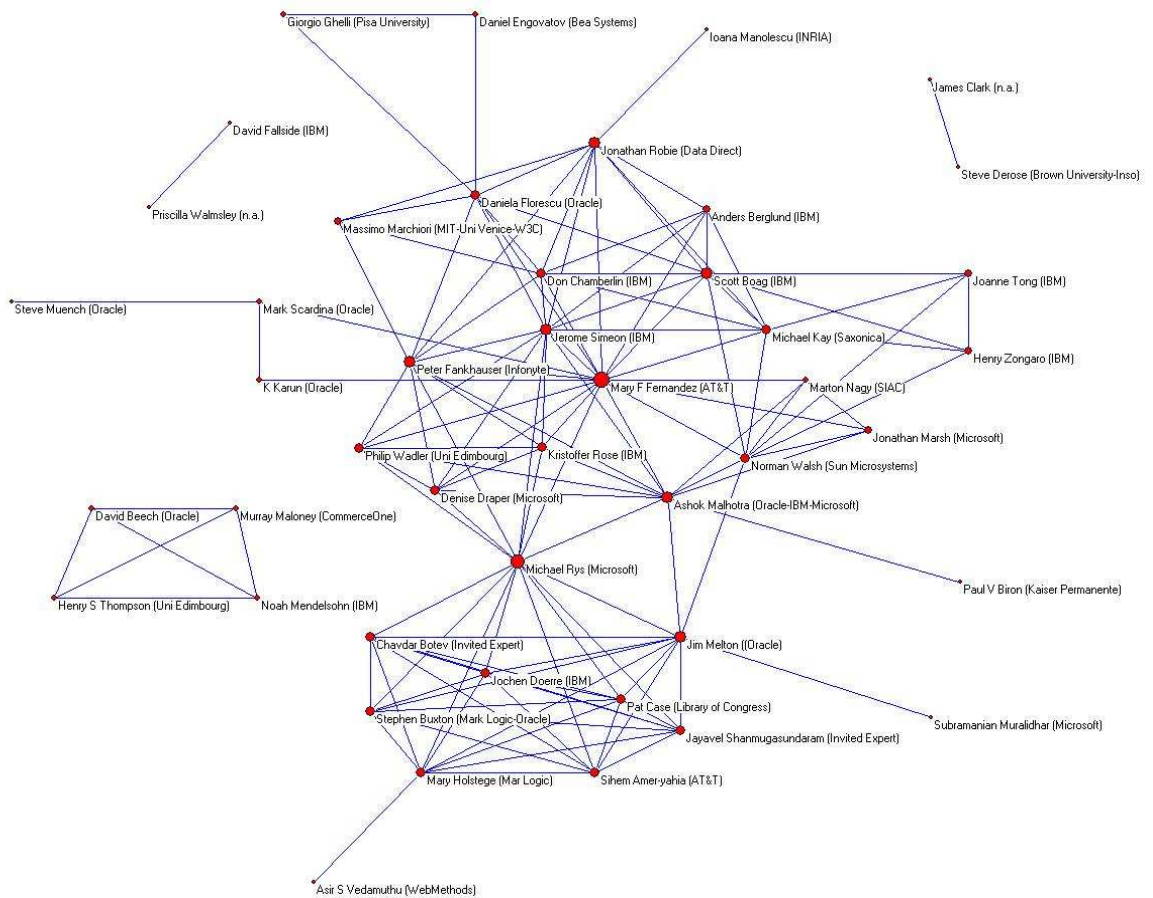

***Graphe 2. Réseau de co-écriture des textes finaux du XQuery au W3C***

**Cercle = auteur de textes**

**La taille des nœuds est relatif au nombre de texte co-écrit.**

La première explication de cette configuration tient dans la différence de moyens consacrés par les firmes à l'activité de normalisation comparés à ceux alloués par les autres types d'acteurs (recherche publique, ONG ou indépendants). Là où les entreprises comme IBM font intervenir jusqu'à 11 experts sur la période considérée, les autres types d'acteurs n'en présentent au mieux que 2. Par ailleurs, l'activité d'expertise au sein du W3C représente pour les chercheurs publics, un investissement coûteux en temps qui vient se superposer à leur activité de recherche, alors que les experts des firmes consacrent généralement l'essentiel de leur activité aux aspects traités par le standard. L'investissement des entreprises dans le processus de normalisation du W3C apparaît donc bien supérieur à celui consentit par les centres de recherche publics et les ONG. L'investissement des « indépendants » est quant à lui beaucoup plus aléatoire, soumis à leur trajectoire individuelle et à leur capacité de s'immerger durablement dans une activité chronophage (David, Rullani, 2008).

Contre toute attente l'orientation *Open Source* du W3C ne joue donc pas comme facteur inhibant à l'égard des firmes. Celles-ci investissent au contraire massivement cette arène et en contrôle effectivement les *outputs*. Quels intérêts ont-elles cependant à participer à des standards sur lesquels elles ne peuvent revendiquer aucun brevet ? Le paradoxe n'est qu'apparent dès que l'on se rappelle que l'informatique supporte plusieurs *business model*. Pour simplifier, on peut considérer qu'il en existe deux principaux. Le premier porté par Microsoft repose sur le modèle classique de la propriété intellectuelle, basé sur la vente de logiciels « prêt à l'emploi » destiné au plus grand nombre. Le second est celui développé notamment par IBM, depuis que la société abandonna la production de *hardware* dans les années 1980, et consiste non pas à vendre des logiciels mais des services informatiques se traduisant par l'élaboration de produits spécifiques pour chaque client. Dans son modèle, IBM n'a que faire de la propriété intellectuelle sur les logiciels, au contraire celle-ci peut rogner ses marges si la société est appelée à avoir recours aux logiciels de Microsoft par exemple. Son intérêt est clairement de travailler dans un monde où la propriété intellectuelle est réduite au minimum, voir n'existe pas, c'est-à-dire dans un monde *Open Source*.

Pour autant que le XML constitue aujourd'hui le nouveau format des logiciels de bureautique et qu'il est sans doute appelé à être le nouveau langage du Web alors que les applications réseaux tendent à se démultiplier (cf. développement récent du *cloud computing*), le lecteur peut à présent saisir tout l'enjeu que représentait la normalisation du XML pour les entreprises.

L'histoire de la standardisation du XML est étroitement liée à l'une de ses applications, celle des logiciels de bureautiques comme *MS Office* de Microsoft ou *OpenOffice* de Sun Microsystem. Mais d'autres entreprises comme IBM ou Oracle se sont aussi intéressées de près à cette technologie et ont comme nous l'avons vu au sein du W3C activement participé à sa standardisation.

### *Le changement d'arène de Sun et IBM vers le consortium Oasis, l'Union Européenne puis ISO*

Jusqu'aux années 2000, Microsoft avait développé, pour sa suite *Office*, un modèle propriétaire de format de fichier qui s'était progressivement imposé comme standard *de facto*. En effet, la position hégémonique de Microsoft sur le marché de la bureautique contraignait ses éventuels concurrents à développer des offres compatibles avec ses produits, au risque de voir leurs logiciels incapables de communiquer avec les produits Microsoft, c'est-à-dire la quasi-totalité du reste du monde. Tous ceux qui, à cette époque, ont connu les problèmes récurrents de conversion de fichiers entre les produits Microsoft et les autres comprendront de quoi il est question. Mais en refusant de communiquer les codes sources de ses logiciels, Microsoft limitait singulièrement cette compatibilité, donc la possibilité même de concurrence, et renforçait ainsi sa position hégémonique. Le XML allait changer la donne. A la fin des années 1990, les grandes firmes d'informatiques ont commencé à développer pour leur propre compte des applications intégrant le XML, alors que la technologie était encore à ses balbutiements et qu'elles concouraient dans le même temps à son élaboration au sein du W3C. Microsoft introduisit pour la première fois le format XML sur *Office 2000*, tandis que Sun Microsytem développait concurremment sa suite OpenOffice XML. Mais alors que Microsoft fidèle à sa tradition maintenait le secret sur ses codes sources, Sun Microsytem proposa, en 2002, sa solution au Consortium OASIS[3] afin d'en faire un standard ouvert basé sur le libre accès aux codes sources. Sa démarche rencontra un succès mitigé et les travaux du consortium n'avancèrent guère en ce sens. Le rapport au temps entre les tenants de l'Open Source (Sun, IBM) d'une part et Microsoft d'autre part était radicalement différent. Alors que les premiers pouvaient souhaiter l'adoption rapide d'un standard ouvert sur le XML, la firme de Redmond avait toutes les raisons de le redouter et d'en repousser l'échéance. Ainsi que nous l'avons indiqué, le business model de Sun et IBM orienté sur les services informatiques (et les réseaux pour Sun) s'accommodait très bien d'un standard ouvert, la différence se faisant sur le service non sur le logiciel en tant que tel. Pour Microsoft c'était l'inverse et même si la firme s'était dès cette époque résolument engagée vers les services informatiques, la transformation complète de l'entreprise aurait pris plusieurs années. Quelle que soit la configuration Microsoft avait besoin de temps avant que ne s'impose un standard du XML. Sun et IBM eurent plus de succès au sein du W3C. Leur implication nettement supérieure à celle de Microsoft dans l'écriture des textes sur le XML montre leur intérêt pour l'émergence d'un standard, chose qui fut faite en février 2004 avec la publication de la première recommandation du W3C sur le XML. Quelques mois plus tard l'Union européenne demandait à OASIS d'avancer sur la standardisation du XML, y compris d'envisager une soumission à l'ISO et à Microsoft de soumettre ses propres développements à un organisme de standardisation international. Dès lors les évènements s'accélérèrent. En janvier 2005, l'Open Office XML format est renommé par OASIS Open Document Format (ODF) afin de détacher la norme de son initiateur Sun. En mai, l'ODF qui reprend pour l'essentiel les spécifications du W3C est adopté comme standard par OASIS et Sun publie en octobre une version *d'Open Office* se revendiquant entièrement de l'ODF. En novembre, le consortium soumet l'ODF à la procédure *Publicly Available Specification* de l'ISO. Le même mois, Microsoft sollicite l'Ecma[4] pour faire reconnaître ses propres développements sur le XML comme standard. Le péril pour Microsoft était alors très grand. Si l'ODF s'imposait comme unique standard du XML appliqué à la bureautique, alors la firme de Redmond aurait été contrainte de l'intégrer à ses logiciels et chose impensable devenir « suiveur » de Sun. En se résignant à promouvoir un standard alternatif, Microsoft acceptait la fin de son business model puisque les codes sources seraient de toute façon accessibles. Mais la société gagnait du temps. La coexistence de deux standards maintiendrait les défauts de compatibilité entre les deux systèmes, donc, pour quelques temps encore, la position hégémonique de Micorosoft. Mais l'adoption de l'ODF par l'ISO en mai 2006, allait accentuer la pression sur le leader mondial. A la différence des autres standards, les normes ISO parce que reprises par les agences nationales ont un degré de contrainte généralement supérieur aux normes de consortium. En faisant de l'ODF une norme ISO, les tenants de l'Open Source obligeaient quasiment Microsoft à abandonner ses développements sur le XML et à reprogrammer Office, à moins que la firme ne parvienne, chose incongrue à plus d'un observateur, à faire de son propre format une deuxième norme ISO. Acculée à challenger le standard, c'est pourtant ce qu'entreprit Microsoft en 2006. La dynamique de changement d'arènes entraînait donc non plus seulement les leaders de l'industrie basée sur l'open source au sein du W3C, mais bien l'ensemble des géants du secteur, dans une bataille de tranchée impliquant de nouvelles ressources et générant des coûts importants.

---

[3] Organization for the Advancement of Structured Information Standards. Il s'agit d'une arène privée plus spécifiquement dédiée à la normalisation des applications informatiques et résolument orientée vers la promotion des standards ouverts. La plus par des grands noms de l'informatique en sont membres y compris Microsoft.

### 4.2. Les ressources mobilisées dans les changements d'arènes

Changer d'arène suppose pour les firmes de convaincre de nouveaux interlocuteurs de la portée pour les producteurs et de l'utilité pour les consommateurs des solutions qu'ils préconisent. Ce travail suppose l'enrôlement progressif de promoteurs du standard et la construction de rapports de force pouvant conduire, au besoin, à des stratégies de retournement des opposants.

---

[4] Initialement European Computer Manufacturer Association. Fondée en 1961 à l'instigation notamment d'IBM, l'Ecma regroupe à l'origine les principaux fabricants européens d'ordinateurs. Elle est dès les années 1960 partie prenante du JT1 ISO/CEI. En 1994, l'association change de nom, Ecma International – European association for standardizing information and communication systems, afin de refléter son internationalisation.

### *L'enrôlement de partenaires dans la promotion du standard*

Le processus d'enrôlement de promoteurs du standard proposé est un processus relativement long et coûteux, puisqu'il suppose de rendre accessible à des partenaires industriels ou commerciaux des solutions techniques qui sont généralement très complexes, et de les engager dans une mobilisation dont l'issue est incertaine, et qui peut aboutir à des rapports de force potentiellement difficiles à soutenir.

Pour Sun, le plein engagement du consortium Oasis en faveur de l'ODF n'aboutit qu'au terme de ce processus, quand des acteurs comme IBM ou l'Union européenne s'engagent publiquement sur la solution définie. L'engagement de la Commission Européenne, qui mène depuis longtemps une politique de sanction de Microsoft pour abus de position dominante sur de multiples dossiers, est ici essentiel pour deux raisons. D'abord, c'est cet engagement et l'incitation à développer un standard ISO qui accélère le mouvement de *shifting*. Ensuite, la perspective d'un débouché rapide sur le marché européen facilite l'engagement d'OASIS.

Pour Microsoft, l'objectif de challenger le standard consiste dès lors à mobiliser de nouveaux alliés, en enrôlant d'autres partenaires ou consortiums en faveur de l'OOXML – en l'occurrence l'ECMA. Ces processus ont pour effet de construire un rapport de forces entre des coalitions industrielles menées par des leaders capables de soutenir politiquement et financièrement la compétition. Ces rapports de force peuvent d'ailleurs être internes au consortium auquel appartiennent deux grandes firmes rivales.

En décembre 2006, par exemple, l'Ecma reconnaît le format de Microsoft OpenXML comme standard par 20 voix pour et une contre, celle d'IBM. L'opposition d'IBM, qui avait annoncé dès le succès de l'ODF à l'ISO son intégration dans sa suite *Lotus Symphony*, est à peu près équivalente à celle de Microsoft au moment de la standardisation ODF. Ces oppositions ont pour effet de faire basculer les contraintes argumentatives de la simple controverse technique appuyée sur des contraintes procédurales, à des véritables polémiques qui débordent les arènes de standardisation. Microsoft accusera ainsi son homologue d'utiliser « la standardisation pour limiter les choix du marché dans un but commercial, et cela sans prendre en compte l'impact négatif sur les choix des consommateurs et l'innovation technologique »[5]. Voir Microsoft défendre le libre choix des consommateurs est pour le moins plaisant, mais cela n'en souligne que mieux l'état d'urgence qui régnait alors dans la société, et la nécessité pour la firme de Redmond de désigner l'adversaire pour mobiliser les indécis. En effet, le succès à l'Ecma ne suffisait pas à faire de l'OOXML un standard aussi fort que l'ODF qui pouvait s'imposer sur le marché sur la base d'une certification ISO, tout particulièrement auprès des administrations publiques européennes grosses consommatrices de bureautique. Il fallait que l'Open XML soit lui aussi un standard ISO. L'urgence pour Microsoft était telle que la firme recourut à la procédure dite du *Fast Track*, qui permet l'adoption d'un standard en quelques mois, (c'est-à-dire sans examens techniques approfondis) s'il est approuvé par 66% des agences nationales et refusé par moins de 25%. La démarche de Microsoft suscita de vives critiques notamment de la part de la communauté des informaticiens qui s'interrogeait sur la nouvelle conversion de Microsoft à l'Open Source. De leur côté, les agences nationales étaient un certain nombre à se montrer sceptiques sur l'opportunité d'adopter un nouveau standard qui ne viendrait pas se substituer au précédant mais coexister avec lui.

---

[5] Tom Robertson et Jean Paoli, "Interoperability, Choice and Open XML", edited on Microsoft Web site, 2007, February 14.
http://www.microsoft.com/interop/letters/choice.mspx

Enrôler des promoteurs suppose donc un processus de débordement des partenariats industriels classiques, et un véritable travail de politisation des débats. Cela est particulièrement vrai dans le cadre des consultations organisées par les agences nationales votant à ISO, où la polémique et le rapport de forces n'ont cessé d'enfler entre 2006 et 2008, suscitant la création de blogs, la multiplication de campagnes téléphoniques, etc. Ce processus de débordement s'explique par le fait que la standardisation à ISO, largement décentralisée dans des agences nationales, multiplie les fronts, avec des registres d'argumentation et de vote (variation du poids de l'administration et des consommateurs) qui ne sont pas stabilisés[6], et pâtit aussi, du fait de la vocation générale des agences, d'un niveau d'expertise moins élevé, ou moins homogène.

Etant donnée cette grande variation des procédures et des règles, et cette moindre homogénéité de l'expertise technique, les firmes peuvent davantage jouer des asymétries d'information à ISO que dans des arènes comme le W3C ou OASIS. Dans le cadre d'une stratégie de changement d'arène, on ne s'étonnera donc pas qu'elles essaient de maximiser leurs chances de réussite en essayant de construire des majorités *ad hoc*. Le cas des votes à ISO sur le standard OOXML illustre bien ce processus. Comme l'indique bien la comparaison du résultat des votes sur l'ODF et sur l'OOXML (Annexe 1), Microsoft a en effet fortement mobilisé des agences nationales qui ne participent pas habituellement à ce type de débats, et qui ont donc pris le statut de *New Comers*, pour voter sur l'OOXML.

---

[6] Sur ce point, voir Daudigeos (2007).

**Tableau 2. Votes à ISO sur l'ODF (2006)**

| P members | Vote |
|---|---|
| Australia | Approval |
| Belgium | Approval |
| Canada | Approval |
| China | Approval |
| Czech Republic | Approval |
| Denmark | |
| Finland | Approval |
| France | Approval |
| Germany | Approval |
| Iran | Approval |
| Ireland | Approval |
| Italy | Approval |
| Japan | Approval |
| Kenya | Approval |
| Korea | Approval |
| Malaysia | Approval |
| Netherlands | Approval |
| New Zealand | Approval |
| Norway | Approval |
| Saudi Arabia | Abstention |
| Singapore | Approval |
| Slovenia | |
| South Africa | Abstention |
| Spain | Approval |
| Switzerland | Approval |
| United Kingdom | Approval |
| USA | Approval |

| O members | Vote |
|---|---|
| Egypt | Approval |
| Hungary Out | Approval |
| India P | Approval |
| Israel | Approval |
| Luxembourg | Approval |
| Poland | Approval |
| Romania | Approval |
| Sri Lanka | Approval |
| Sweden | Approval |

**Tableau 3. Votes à ISO sur l'OOXML (2007 et 2008)**

| Pays (P-Members) | 1er vote (Source opensourcener) | 2ème vote (Source Open Malaysia) |
|---|---|---|
| Australie (P) | Abstention | Abstention |
| Azerbaijan (N) | Oui | Oui |
| Belgique (P) | Abstention | Abstention |
| Canada (P) | Non | Non |
| Chine (P) | Non | Non |
| Côte d'Ivoire (N) | Oui | Oui |
| Chypre (N) | Oui | Oui |
| République Tchèque (P) | Non | Oui |
| Danemark (P) | Non | Oui |
| Equateur (N) | Non | Non |
| Finlande (P) | Abstention | Oui |
| France (P) | Non | Abstention |
| Allemagne (P) | Oui | Oui |
| Inde (O) | Non | Non |
| Iran (P) | Non | Non |
| Irlande (P) | Non | Oui |
| Italie (P) | Abstention | Abstention |
| Jamaique (N) | Oui | Oui |
| Japon (P) | Non | Oui |
| Kazakhstan (N) | Oui | Oui |
| Kenya (P) | Oui | Abstention |
| Corée (P) | Non | Oui |
| Liban (N) | Oui | Oui |
| Malaysie (P) | Abstention | Abstention |
| Malte (N) | Oui | Oui |
| Pays-Bas (P) | Abstention | Abstention |
| Nouvelle Zélande (P) | Non | Non |
| Norvège (P) | Non | Oui |
| Pakistan (N) | Oui | Oui |
| Arabie Saoudite (P) | Oui | Oui |
| Singapore (P) | Oui | Oui |
| Slovénie (P) | Abstention | Oui |
| Afrique du Sud (P) | Non | Non |
| Espagne (P) | Abstention | Abstention |
| Suisse (P) | Oui | Oui |
| Trinidad et Tobago (N) | Abstention | Oui |
| Turquie (N) | Oui | Abstention |
| Royaume-Uni (P) | Non | Oui |
| Uruguay (N) | Oui | Oui |
| USA (P) | Oui | Oui |
| Venezuela (N) | Oui | Non |

(P) = P-member à l'ODF ; (O) = O-Member à l'ODF ; (N = *Newcomer* (pas présent à l'ODF)

| Pays (O et autres) | 1$^{er}$ vote | 2$^{ème}$ vote |
|---|---|---|
| Arménie | Oui | Oui |
| Argentine | Abstention | Abstention |
| Autriche | Oui | Oui |
| Bengladesh (S) | Oui | Oui |
| Barbade (S) | Oui | Oui |
| Biélorussie | Oui | Oui |
| Bosnie Herzégovie (S) | Oui | Oui |
| Brésil | Non | Non |
| Bulgarie | Oui | Oui |
| Chili | Abstention | Abstention |
| Congo (S) | Oui | Oui |
| Colombie | Oui | Oui |
| Costa Rica | Oui | Oui |
| Croatie | Oui | Oui |
| Cuba | Oui | Non |
| Egypte (O) | Oui | Oui |
| Fiji (S) | Oui | Oui |
| Ghana (S) | Oui | Oui |
| Grèce | Oui | Oui |
| Israel (O) | Abstention | Oui |
| Jordanie (S) | Oui | Oui |
| Kowait (S) | Oui | Oui |
| Luxembourg (O) | Abstention | Abstention |
| Ile Maurice (S) | Abstention | Oui |
| Mexique | Abstention | Oui |
| Maroc | Oui | Oui |
| Nigéria (S) | Oui | Oui |
| Panama (S) | Oui | Oui |
| Pérou | Abstention | Oui |
| Philippines | Non | Oui |
| Pologne (O) | Oui | Oui |
| Portugal | Oui | Oui |
| Qatar (S) | Oui | Oui |
| Roumanie (O) | Oui | Oui |
| Russie | Oui | Abstention |
| Sri Lanka (O) | Oui | Abstention |
| Serbie | Oui | Oui |
| Syrie (S) | Oui | Oui |
| Tanzanie (S) | Oui | Oui |
| Thailande | Non | Oui |
| Tunisie | Oui | Oui |
| Emirats Arabes Unis (S) | Oui | Oui |
| Ukraine | Oui | Oui |
| Ouzbékistan (S) | Oui | Oui |
| Vietnam | Abstention | Abstention |
| Zymbabwe (S) | Abstention | Abstention |

(S) = Sans statut ; (O) = O-member à l'ODF

Comme l'indiquent ces tableaux, l'ODF a fait l'objet d'une quasi-unanimité[7]. Les 27 P-members s'étant prononcés sur l'ODF sont restés des P-members lors du vote sur l'XML . 6 O-members sur 9 sont restés des O-members. Un O-member est devenu un P-member (Inde), et 1 est parti (Hongrie). Donc, parmi 41 les P-members s'étant prononcés lors du vote sur l'XML, 27 avaient déjà participé au vote sur l'ODF entant que P-members et 1 entant que O-member. Il y avait donc, lors du vote sur l'XML, 13 *newcomers* parmi les P-members. En ce qui concerne les O-members, ils étaient 9 pendant la procédure sur l'ODF. Sur ces 9, 7 sont restés des O-members durant le vote sur le XML et 1 est devenu P-member (Inde). Cela dit, durant le vote sur le XML le nombre de O-members s'élevait à 30. Si à cela on ajoute ceux qui n'avaient aucun statut (ni O ni P), il y avait pour le XML 40 *newcomers* parmi les non p-members. Donc, toutes catégories confondues, sur les 87 participants au vote sur le XML, 53 n'avaient pas participé au vote sur l'ODF, ce qui est considérable. Il s'agit d'un constat intéressant dans le sens où il témoigne de l'importance accordée à la question.

Comme on l'a dit, l'OOXML a tout d'abord été refusé. Il y a donc eu deux rounds. Notons d'emblée qu'entre ces deux rounds, un nombre non négligeable de pays qui avaient pris des positions pro-ODF ont finalement voté oui sur l'OOXML. Si l'on s'intéresse aux P-members, sur 41 pays, 26 ont maintenu un vote positif et 14 ont changé d'avis. Sur ces 14, 7 sont passés de non à oui (République Tchèque, Danemark, Irlande, Japon, Corée, Norvège, Royaume-Uni), 3 de abstention à oui (Finlande, Slovénie, Trinidad et Tobago), 1 de non à abstention (France), 2 de oui à abstention (Kenya et Turquie) et 1 de oui à non (Venezuela). Il serait intéressant de se pencher plus en détail sur les raisons de ces changements. Sur ces 41 pays, 13 étaient des *Newcomers*[8] (Azerbaijan, Côte d'Ivoire, Chypre, Equateur, Jamaique, Kazakhstan, Liban, Malte, Pakistan, Trinidad et Tobago, Turquie, Uruguay, Venezuela). Sur ces 13, 9 ont maintenu un vote positif d'un round à l'autre (Azerbaijan, Côte d'Ivoire, Chypre, Jamaique, Kazakhstan, Liban, Malte, Pakistan, Uruguay) Et un a maintenu un vote négatif (Equateur).  Plus généralement, 14 pays ont maintenu un vote positif (Azerbaijan, Arabie Saoudite, Côte d'Ivoire, Chypre, Jamaique, Liban, Malte, Pakistan, Uruguay, Allemagne, Kazakhstan, Singapore, Suisse, USA). Sur ces, 14, comme nous l'avons vu avant, 9 étaient des *Newcomers*. 7 pays ont maintenu un vote négatif. Au final, en ce qui concerne les p-members, nous avons 24 oui, 8 non et 9 abstentions. Selon le premier critère, à savoir qu'il faut le vote positif de plus de 66% des p-members, la norme passe avec 24/32 des voix (les abstentions ont été enlevées). Mais en ce qui concerne le deuxième critère, selon lequel le projet ne doit pas compter plus de 25% de votes négatifs sur *l'ensemble* des votes (donc toutes catégories confondues), nous devons également prendre en compte le tableau suivant.

---

[7] Parmi les p-members, deux pays se sont abstenus (Arabie Saoudite et Afrique du Sud) et deux n'ont pas voté (Danemark et Slovénie). Tous les autres ont approuvé avec ou sans réserves, le plus souvent sans.

[8] Par là nous entendons qu'ils n'étaient pas là pendant la procédure sur l'ODF.

Sur 46 pays non p-members, 30 ont le statut d'observateur et ont donc participé aux négociations et aux discussions. Les 16 autres ont le droit de vote, mais n'ont pas eu le droit de participer pas au processus en amont. 31 pays ont conservé un vote positif. Parmi ceux-ci, 14 n'ont pas le statut d'observateur (en tout ils sont 16). C'est un point intéressant car ces pays n'ont pas participé au processus de négociation et de discussion mais ils se sont prononcés sans équivoque en faveur de la norme (15 sur 16). Même l'Ile Maurice, qui a changé son vote, est passée d'abstention à oui. Il est difficile de dire s'il y a un lien mais les chiffres sont intéressants. Un seul pays a conservé un vote négatif (Brésil).

En tout 16 pays (toutes catégories confondues) se sont abstenus. Cela réduit le nombre de participants à 71. Sur ces 71, 9 se sont finalement opposés ce qui fait environ 13% ce qui fait passer la norme.

Comme on le voit après ce commentaire des votes, construire des majorités suppose d'enrôler des agences nationales qui ne sont pas a priori intéressées par les enjeux de la standardisation, ou dont l'activité dans le secteur reste modeste. Mais cet enrôlement ne suffit pas toujours à construire les majorités. Entre les deux rounds, il a fallu pour Microsoft déployer des moyens considérables pour retourner les opposants, afin de limiter le nombre de suffrages négatifs.

### *Le retournement des opposants*

Les cas de la Suède et de la France sont ici éclairants pour comprendre l'intensité du travail effectué pour retourner les votes des agences nationales a priori peu disposées à l'égard de l'OOXML.

*Le cas suédois*

La Suède s'est retrouvée sous les feux des projecteurs pour une affaire de corruption. La Suède est représentée à l'ISO par le SSI (Swedish Standards Institute). La Suède a, dans un premier temps, voté en faveur de l'adoption d'Open Xml comme norme internationale mais a ensuite décidé d'invalider son vote à cause d'un vice de procédure. Le vote favorable de la Suède a éveillé les soupçons parce que les débats menés au sein du SSI semblaient être plutôt hostiles à Open Xml[9]. Après que 23 sociétés, pour la plupart proches de Microsoft, s'y soient inscrites (l'inscription coûte, selon les sources entre 1800 et 2200 euros) les débats ont changé d'orientation.[10] En effet le vote s'est clos sur un « oui » écrasant : 25 « oui », 6 « non » et 3 « abstention ». Ce qui a étonné les commentateurs, c'est que « la plupart des sociétés venues récemment au SSI n'avaient jamais montré de signes particuliers d'intérêt pour la question globale des standards. Face à un tel enchaînement des faits, les premiers commentaires à vif pensent sans détour que Microsoft a acheté le vote de la Suède»[11]. En effet, peu de temps après les faits, le magazine *Computer Sweden*[12] révèle que « la filiale suédoise de Microsoft avait promis quelque « bonus marketing » à ses partenaires s'ils votaient en faveur d'Open XML »[13]. Cela dit, comme le rapporte Yves Grandmontagne dans son article « Norme ISO d'Open XML: Microsoft gagne une manche, sans plus... »[14], Microsoft s'est défendu en affirmant que ce genre de pratiques étaient tout à fait légales et que tout s'était fait dans le respect des règles. De plus, si elle a reconnu qu'un de ses employés avait promis des primes, elle ne se considère pas comme responsable de ses actes. Tom Robertson, manager general pour l'interopérabilité et les standards à Microsoft a, en effet, déclaré : « We had a situation where an employee sent a communication via e-mail that was inconsistent with our corporate policy »[15]. C'est également ce qu'affirme Jason Matusow (Microsoft's senior director for intellectual property and interoperability) dans son blog. Ce dernier donne sa version des faits en disant qu'un employé[16] a envoyé un mail à deux partenaires

---

[9] « Sweden: heavily stacked "yes" vote was retracted for a technicality », in http://www.noooxml.org/irregularities, voir document Suède_Article_SwedenVoteRetracted

[10] La liste des sociétés inscrites est rapportée dans HAVERBALD, Kim, « Microsoft buys the Swedish vote on OOXML », in http://www.os2world.com/content/view/14868/2/, voir document Suède_Article_MicrosoftBuySwedishVote, les affiliations des différentes sociétés sont également données. Nous n'avons, cela dit, pas confirmé l'information. Voir également le document Suède_Article_MicrosoftAcheteVotes

[11] « Open XML : une standardisation difficile et sous influence, une vision particulière de l'ouverture », in http://www.pcinpact.com/actu/news/38517-open-xml-microsoft-iso-standardisation-votes.htm, voir document Suède_Article_OpenXml_StandardDifficile

[12] Voir document Suède_Article_ComputerSweden

[13] CHAUSSON, Cyrille, « Open XML : La Suède en route vers l'abstention », Edition du 31/08/2007, in http://www.lemondeinformatique.fr/actualites/lire-open-xml-la-suede-en-route-vers-l-abstention-23815.html, voir document Suède_Article_SuèdeEnRouteAbstention, voir également le document Suède_Article_MicrosoftForcedVoteYes

[14] GRANDMONTAGNE, Yves, « Norme ISO d'Open XML: Microsoft gagne une manche, sans plus... », 02-09-2007, in http://www.silicon.fr/articles/printView/fr/silicon/news/2007/09/02/man-uvres-de-microsoft-aller-l, voir document Suède_Article_MicrosoftGagneSansPlus

[15] « Sweden's SIS Declares OOXML Vote Invalid - Will Change Vote from Yes to Abstain », Thursday, August 30 2007 @ 03:51 PM EDT, in http://www.groklaw.net/articlebasic.php?story=20070830155109351, voir document Suède_Article_SwedenVoteInvalid

[16] Microsoft a, en effet, annoncé qu'il s'agissait de l'œuvre d'un seul employé. Or il est assez douteux qu'un employé de Microsoft se permette de telles libertés en parlant au nom de la compagnie sans en avoir l'autorité. Il s'agit là de notre avis personnel. Aucune preuve tangible

de Microsoft pour leur signifier que s'ils s'inscrivaient au SSI et votaient en faveur de l'adoption de la norme Open Xml ils recevraient des compensations. Ce même employé, conscient d'agir à l'encontre de la ligne directrice de Microsoft, serait revenu sur son acte et aurait renvoyé un mail pour dire aux deux partenaires en question de ne pas prendre en compte le mail précédemment envoyé. Matusow dit également que lorsque la direction de Microsoft a pris connaissance de la situation ils ont immédiatement averti le SSI pour leur communiquer ce qu'il s'était passé et que cela n'a eu aucun incident sur le vote.[17] Enfin il affirme que beaucoup de parties étaient à la fois des partenaires de Microsoft et d'IBM et que ces derniers auraient également poussé certains partenaires à participer au comité. L'importance de ces controverses montre bien la pression très forte qui pèse sur l'ensemble des acteurs dans ce type de négociations.

Finalement, ne pouvant pas organiser un second vote avant le 2 septembre (date butoir), la Suède s'est donc abstenue.[18]

*Le cas français*

De son côté, la France est passée d'un round à l'autre de « Non avec commentaires » à « abstention avec commentaires ».[19] La France est représentée à ISO par l'AFNOR (Association française de normalisation). Le 10 mai 2007, cet organisme a crée une commission de normalisation spécialisée « pour élaborer par consensus la position française [en ce qui concerne Open Xml] et traiter les commentaires issus de l'enquête probatoire »[20] : la Commission de normalisation Formats de documents révisables (CN FDR). Cette dernière est composée « de l'Etat et de communautés ayant un intérêt direct sur le sujet et qui en feraient la demande »[21].

---

de l'implication de la direction de Microsoft n'est à notre disposition. Nous ne pouvons donc que soulever des pistes.

[17] « Open XML - The Vote in Sweden », in
http://blogs.msdn.com/jasonmatusow/archive/2007/08/29/open-xml-the-vote-in-sweden.aspx, voir document Suède_Article_VoteInSweden

[18] « Sweden's SIS Declares OOXML Vote Invalid - Will Change Vote from Yes to Abstain », Thursday, August 30 2007 @ 03:51 PM EDT, in
http://www.groklaw.net/articlebasic.php?story=20070830155109351, voir document Suède_Article_SwedenVoteInvalid, sur cette page il y a un rapport traduit de ce qu'a déclaré le SSI

[19] Les commentaires sont disponibles sur le document France_ISO_Commentaire_France(AFNOR).doc.

[20] AFNOR, « Projet de norme Office Open XML : la réponse d'AFNOR à la consultation internationale organisée par l'ISO », *Conférence de presse*, Lundi 3 septembre 2007, p. 9, Voir document France_AFNOR_ConférencePresse_03_09_2007

[21] *Idem,* la liste précise des membres est disponible dans le document France_AFNOR_ListeMembres_CNFDR.pdf.

Après l'annonce du vote négatif, avec commentaires, sur le format Office Open XML, un début de polémique a fait surface. En effet, les représentants de Microsoft France (particulièrement Marc Mossé et Bernard Ourghanlian) ont vivement critiqué le déroulement de la réunion de la commission lors d'une conférence téléphonique donnée le soir même de l'annonce de l'avis de l'AFNOR, le 3 septembre 2007 à 19h. Plus précisément, ils ont condamné le manque de consensus (et soutenu que l'abstention était la meilleure solution), la mise en ligne des travaux de la commission de normalisation au fur et à mesure des débats (alors qu'ils étaient sensés être confidentiels), l'importance des pressions extérieures (ainsi qu'un important afflux de votants de dernière minute, critique intéressante car Microsoft a elle-même été accusée d'utiliser ce genre de pratiques), le climat tendu des débats (certains participants auraient été insultés) ainsi que le fait que « l'AFNOR ait publié plus de 120 pages de commentaires, tous en provenance de la même société (IBM mais elle n'est pas nommée), sans que les propositions n'aient été débattues un seul instant entre les divers acteurs de la commission »[22] (fait qui a été démenti par François Legendre, responsable développement et communication pour la normalisation dans le secteur IT de l'Afnor.[23]

Marc Mossé a également insisté sur le fait qu'il fallait prendre en compte l'avis des utilisateurs et « se méfier quand les informaticiens parlent aux informaticiens »[24], surtout que que les utilisateurs étaient généralement favorables à l'adoption d'Open Xml, ce que semble confirmer l'avis de l'Association française des usagers des services de l'administration électronique (Afusae). En effet, selon cette dernière, « le plus positif serait que les deux formats disposent de la normalisation ISO (…). Cela créerait les conditions d'une saine concurrence dont bénéficieraient les utilisateurs »[25]. Ce point a de quoi attirer notre attention car axer sa stratégie sur les utilisateurs est quelque chose de nouveau pour Microsoft.

    Sachant que le sujet était polémique et que sa réponse risquait de provoquer la colère de Microsoft, l'Afnor a immédiatement donné une

---

[22] JAY, Julien, « Open XML&ODF: l'AFNOR demande la convergence », in http://www.neteco.com/79421-interview-afnor-microsoft-openxml-odf.html, voir le document France_Article_OpenXML&ODF_AfnorConvergence

[23] voir le document Article_AfnorsExplique. Pour plus de précisions sur ces points, se référer aux documents France_Article_MicrosoftPilonneAfnor, France_Article_MicrosoftCritiqueDécisionAfnor, France_Article_OpenXML&ODF_AfnorConvergence, France_Article_OOXMLMicrosoftSurpris.

[24] LEMAIRE, Bertrand, « Open XML : Microsoft pilonne l'Afnor », in http://www.cio-online.com/actualites/lire-openxml-microsoft-pilonne-l-afnor-774.html, édition du 03/09/2007, voir document France_Article_MicrosoftPilonneAfnor

[25] GUILLEMIN, Christophe, « Microdoft surpris de la décision de l'Afnor », in http://www.zdnet.fr/actualites/informatique/0,39040745,39372625-2,00.htm, voir document France_Article_OOXMLMicrosoftSurpris.

conférence de presse, afin de justifier son choix. L'Afnor justifie son « non avec commentaires » par le manque de consensus et la nécessité d'une convergence entre ODF et OOXML.[26].

Plusieurs témoignages convergent sur le fait que l'évolution du vote de l'AFNOR à ISO est directement lié à des injonctions politiques, les fonctionnaires de Bercy ayant été invités à se plier à ce choix politique. Si nous n'avons pas accès aux tractations qui se sont nouées en haut, plusieurs acteurs de ce processus témoignent d'un travail de lobbying très intense, développé dans des répertoires aussi variés que les classiques invitations à dîner, le chantage à l'emploi, ou les coups de téléphones persuasifs aux cadres des grandes entreprises françaises.

Le choix de l'AFNOR s'inscrivait en effet dans le contexte de l'adoption du Référentiel général d'interopérabilité (RGI), introduite par l'ordonnance n° 2005-1516 du 8 décembre 2005, qui désigne un travail sur l'interopérabilité des systèmes mis au point par une large collaboration de toutes les administrations sous l'égide de la DGME. Plus précisément, « l'objet du RGI est de fixer les règles techniques permettant d'assurer l'interopérabilité de tout ensemble de moyens destinés à élaborer, traiter, stocker ou transmettre des informations faisant l'objet d'échanges par voie électronique entre autorités administratives et usagers ainsi qu'entre autorités administratives »[27]. Or en ce qui concerne l'échange de documents bureautiques, le RGI recommande, *a priori*, le format ODF, d'autant plus que, comme l'affirme Lemaire, « l'administration s'est dirigée massivement, depuis des années, vers la suite bureautique OpenOffice qui utilise le format OpenDocument, cela pour des raisons de coût et de facilité de déploiement sans gestion onéreuse des licences et surtout sans le versioning au rythme imposé par un éditeur indépendamment de tout besoin métier »[28]. Selon Reynald Fleychaud[29], le chef de service de la DGME aurait affirmé que « Le projet de RGI présenté lors du dernier comité des référentiels du 12

---

[26] Pour plus de détails techniques, Voir :
France_Article_MicrosoftEnRouteCanossa,
France_Article_OpenXML&ODF_AfnorConvergence,
France_Article_OpenXML&ODF_AfnorConvergence2,
France_Article_OpenXML&ODF_AfnorConvergence3,
France_Article_OpenXML&ODF_AfnorConvergence4,
France_AFNOR_ConférencePresse_03_09_2007
[27] MINISTERE DU BUDGET, DES COMPTES PUBLICS ET DE LA FONCTION PUBLIQUE, « Référentiel Général d'Interopérabilité, Interopérabilité Technique, Normes et Recommandations », version de travail 0.98c, voir document
France_Referentiel_General_d_Interoperabilite_-_volet_technique_v0., voir également les documents France_Article_RGI5, France_Article_RGI6, France_Article_RGI7,
France_Article_RGI8, France_Article_RGI9
[28] LEMAIRE, Bertrand, « les dessous du scandale du RGI », in http://www.cio-online.com/actualites/lire-exclusif-les-dessous-du-scandale-du-rgi-1342.html, voir document France_Article_RGI3
[29] Flechaux Reynald, « OpenXML : aussitôt normalisé, aussitôt adoubé par l'Etat »,

octobre 2007 avait été mis en attente, suite à la démarche engagée à l'ISO par l'ECMA concernant le standard OpenXML. Cette démarche ayant maintenant abouti, nous en avons tenu compte et nous souhaitons engager sans délai la démarche de validation du RGI, pour une présentation du projet aux assises du numérique de fin mai 2008 ». L'auteur y voit une preuve du fait que le gouvernement attendait OpenXml.[30] C'est également ce que soutiennent Bertrand Lemaire et Miléna Nemec-Poncik qui, dans leur article « Les dessous de l'adoption du Référentiel d'interopérabilité de l'administration française »[31], font référence à une note récupérée par leurs collègues du CIO (Groupe IT News info) remise par Microsoft au gouvernement français, note qui serait à l'origine du retard de la publication du RGI.[32] Quoi qu'il en soit, comme c'est souligné dans un article de Bertrand Lemaire intitulé « La grande tournée de Microsoft dans les palais de la République »**[33]**, Steve Ballmer, patron de Microsoft, « s'est rendu à l'Assemblée Nationale pour y rencontrer des députés et, mardi 2 octobre, a consacré une autre visite le midi au Premier Ministre François Fillon. Mercredi 3 octobre, c'est les principaux représentants de Microsoft France (le PDG Eric Boustouller, le directeur technique Bernard Ourghanlian, et le chargé des affaires publiques Marc Mossé) qui a rencontré la direction de la DGME (Direction générale de la modernisation de l'Etat) ». Cela dit, si beaucoup de témoignages concordent, il est difficile de réunir autre chose que des indices de ces « opérations de relations publiques ». Notons néanmoins que ces visites ont eu lieu moins de 10 jours avant la réunion du comité RGI où l'affaire fut bloquée.

Le fait que l'Afnor ait changé son vote de non a abstention a ensuite soulevé de nouvelles controverses, mais cette fois du côté des opposants à Microsoft et notamment de l'April (une association de défense du logiciel libre). Cette dernière, dans un communiqué de presse fait le 02 avril 2008, dit que « le vote de l'Afnor ne reflète absolument pas un consensus de la Commission, bien au contraire. Lors de la réunion du 25 mars 2008 destinée à finaliser la position de la commission, tous les membres présents n'était pas un choix acceptable »[34], constat qui est également présent dans une interview de Frédéric Couchet,

---

[30] Voir également le document France_Article_GuiltyPartiesOOXML

[31] LEMAIRE, Bertrand, NEMEC_PONCIK, Miléna, « Les dessous de l'adoption du Référentiel d'interopérabilité de l'administration française », in http://www.reseaux-telecoms.net/actualites/lire-les-dessous-de-l-adoption-du-referentiel-d-interoperabilite-de-l-administration-francaise-18173.html, voir document France_Article_RGI1

[32] Voir également les documents suivants :
France_Article_MicrosoftImposeOOXMLAAdmin, France_Article_RGI2, France_Article_RGI3 (dans lequel la « note de Microsoft », bien que pas publiée est abondamment commentée), France_Article_RGI4

[33] LEMAIRE, Bertrand, « La grande tournée de Microsoft dans les palais de la République », in http://www.cio-online.com/actualites/lire-la-grande-tournee-de-microsoft-dans-les-palais-de-la-republique-847.html, voir document France_Article_TourneeMicrosoftPalaisRepublique

[34] APRIL, « OOXML – la norme adoptée d'avance », *conférence de presse*, Paris, le 02 avril 2008, in http://www.april.org/articles/communiques/pr-20080402.html, voir document Article_APRIL_ConférenceDePresse

fondateur et délégué général de l'April, ayant siégé en son nom à la CN FDR, donc témoin direct.[35] Ce dernier ajoute cependant dans son interview un élément intéressant à savoir le fait que la Direction Générale des Entreprises (DGE) et la Direction Générale de la modernisation de l'Etat (DGME) (qui représentent le gouvernement), « étaient clairement en faveur du non lors de cette réunion »[36]. Or il se trouve que celles-ci ont changé d'avis à la dernière minute (le vendredi 28 mars), ce qui aurait apparemment justifié le changement de vote.

Le cas suédois et le cas français sont deux cas parmi d'autres qui exemplifient l'important travail de retournement des opposants initié par Microsoft et le contexte de mobilisation générale de l'ensemble des acteurs nationaux (concurrents directs, associations d'usagers, militants de l'open source, haute administration). De l'aveu de certains d'entre eux ayant une longue expérience de la standardisation, rarement les échanges ont atteint ce niveau de mobilisation sur des enjeux informatiques.

## 5. Vers une analyse dynamique de la gouvernance des standards

Ce cas d'étude édifiant, et sans doute plus médiatisé que bien d'autres, lève un coin de voile sur la complexité du paysage de la standardisation et de ses approches stratégiques par les firmes[37]. Si les agences internationales ont perdu de leur centralité, en étant de plus en plus des lieux de consécration des standards plutôt que des arènes d'élaboration, elles continuent de peser d'un poids important. Toutefois, l'adoption par l'ISO de deux standards tend dans le même temps à affaiblir la position de cette institution en montrant son incapacité à trancher une querelle économique. Accepter deux standards allait à l'encontre même de la philosophie de l'ISO qui pour une technologie particulière recherche l'universalité et non la pluralité. Pourtant, après un premier échec en 2007, Microsoft est parvenu à faire reconnaître son standard par l'ISO au printemps 2008 avec l'adoption de la norme ISO/IEC DIS 29500, appelée OOXML.

---

[35] « L'Affaire AFNOR: Interview avec Frédéric Couchet d'APRIL sur OOXML en France », in http://www.groklaw.net/article.php?story=20080419005201783#french, voir document France_Itw_FredCouchet.

[36] *Idem*

[37] Pour une présentation de ce type de stratégie par Orange, cf . P. Lucas (VP International standardisation and Industry relationships), *Standardisation landscape & impact of Internet in the telecom industry*, Journée d'étude internationale 'Régulation de l'internet : des standards techniques aux normes sociales' Chaire Orange Economie numérique et innovation et Vox Internet, Ecole Polytechnique de Paris, Paris, 31 mars 2009, diapos en ligne sur le site dédié de Vox Internet.

Le succès de Microsoft est cependant une victoire à la Pyrrhus. La firme a certes permis à la version Office 2007 de pouvoir être commercialisée dans de bonnes conditions[38], mais a dû dans le même temps abandonner le secret qu'elle entretenait sur ses codes sources. Au delà de la victoire de tel ou tel standard, c'est le business model de la bureautique d'entreprise pour les prochaines années qui s'est joué avec le XML. De ce point de vue, ce sont les services informatiques, marché sur lequel IBM est le leader mondial, qui ont remporté une victoire décisive. L'enseignement majeur de ce processus de standardisation reste cependant la très nette domination des firmes et les usages stratégiques qu'elles font des arènes de normalisation dans le contexte plus large des luttes compétitives auxquelles elles se livrent.

Comme le souligne Sheila Jasanoff (2005, p. 214): " A un moment où une très grande majorité de décisions publiques implique des éléments non négligeables d'analyse technique, tout changement dans les positions relatives du jugement scientifique et du jugement politique produit un déplacement de l'exercice du pouvoir, avec des conséquences en termes de participation, de débat et de responsabilité. Pas moins que dans les années soixante, quand le politiste de Yale Robert Dahl proposait le titre de son étude séminale de la démocratie, la question reste « Qui gouverne ? » (Dahl 1961). Une différence néanmoins, tient au fait que les modes de décision techniques sont désormais beaucoup plus continuellement et visiblement une part du terrain de jeu politique ».

Poser la question de la gouvernance de la standardisation ne revient donc pas à dépolitiser l'analyse de la décision technique, mais au contraire à la réintégrer beaucoup plus franchement dans une perspective d'économie politique qui ne définisse pas a priori les acteurs les plus légitimes, et qui n'opère pas plus de distinction implicite entre les bons et les méchants dans la compétition économique, comme c'est trop souvent le cas sur ces questions. Répondre à la question « Qui gouverne la standardisation ? » suppose de s'intéresser aux conditions d'échange et aux modes de structuration de la gouvernance sur l'ensemble de la séquence pertinente. Dans le cas présent, nous voyons de manière évidente le passage d'une structure de gouvernance transnationale empruntant certains traits à la gouvernance en réseau (spécificité des actifs humains au W3C, complexité des tâches, réputation et légitimité professionnelle des experts), à une structure beaucoup plus typique de la gouvernance transnationale privée, avec un contexte de redéfinition de l'organisation des marchés (émergence de l'open source), des mécanismes de coordination lourds (organisation des rounds ISO), une compétition oligopolistique (Sun, IBM, Oracle, Microsoft), la dépendance vis-à-vis de chaînes de valeur globales (intégration dans Office), le développement de rapports de force au prix de plusieurs dizaines de millions de dollars, la mise en cause de la crédibilité des firmes, etc.

---

[38] Elle a néanmoins encore une fois été condamnée par la Commission européenne à plus de 220 millions d'euros d'amende sur ce point en août 2009.

Dans ce travail de qualification de la gouvernance, il faut insister sur l'interdépendance entre les arènes de normalisation et la capacité des firmes à contrôler ce travail par les stratégies de changement d'arène. C'est pourquoi il faut aller au-delà des analyses réalisées à partir de monographies d'arènes de standardisation, qui, si elles ont leur utilité, donnent souvent une vision trop partielle ou trop synchronique de ce travail. Analyser ces processus à partir d'analyses diachroniques ou longitudinales des réseaux d'acteurs et non des organisations formelles permet de dynamiser l'analyse et d'approcher la formidable complexité de ces processus, pour en faire ressortir ensuite les dynamiques économiques et sociales les plus saillantes.

C'est peut-être en développant ce type d'analyse dynamique que l'économie politique internationale peut contribuer à éclairer les enjeux du moment, et réussir à prendre pied avec force dans les débats économiques, sans trop de complexes vis-à-vis des modélisations plus ou moins sophistiquées, qui, de par leur caractère statique ou linéaire, passent très régulièrement à côté de la complexité des jeux sur des séquences longues.

## 6. Bibliographie

## 6. WEBSITES

QizX Open: a free-source Xquery Engine. Available at http://www.axyana.com/qizxopen/

Active XML reference: http://www.axml.net/

http://www.afnor.org

http://www.fondonorma.org

http://www.gobiernoenlinea.ve

http://www.iso.org

http://www.ecma-international.org

http://www.oasis-open.org

http://www.blog.bureado.com.ve http://fr.wikipédia.org/wiki/Format_ouvert
http://aporrea.org